\documentclass[namedreferences,hyperref,optionalrh]{spr-sola}
\usepackage{graphicx}        % For eps figures, newer & more powerfull
\usepackage{color}           % For color text: \color command
\usepackage{amsmath}
\usepackage{amsfonts}
\usepackage{siunitx}
\renewcommand{\vec}[1]{{\mathbfit #1}}

\chardef\us=`\_

\begin{document}

\begin{frontmatter}
\title{Linear analysis of shear-flow instabilities in a prominence-corona interface with  ambipolar diffusion}

\author[addressref={aff1,aff2},email={llorenc.melis@uib.cat}]{\inits{L.}\fnm{Lloren\c{c}}~\snm{Melis}}
\author[addressref={aff1,aff2},email={roberto.soler@uib.es}]{\inits{R.}\fnm{Roberto}~\snm{Soler}}
%\author{\inits{}\fnm{}~\lnm{}\orcid{}}
%   NOTE:  Just one corresponding author [corref]
\address[id=aff1]{Departament de Física, Universitat de les Illes Balears, E-07122, Palma de Mallorca, Spain}
\address[id=aff2]{Institut d'Aplicacions Computacionals de Codi Comunitari (IAC3), Universitat de les Illes Balears, E-07122, Palma de Mallorca, Spain}

\runningauthor{Lloren\c{c} Melis \& Roberto Soler}
\runningtitle{Shear-flow instabilities  with  ambipolar diffusion}

\begin{abstract}
 Observations have shown the  presence of  the Kelvin-Helmholtz instability (KHi) in solar prominences. Effects due to partial ionization of the prominence plasma may influence the KHi onset. We study the triggering of the KHi in an interface model that consists of a partially ionized prominence region and a fully ionized coronal region, with a uniform magnetic field parallel to the interface. There is a longitudinal flow in the prominence region. The plasma is compressible and the role of ambipolar diffusion, which accounts for  collisions between charges and neutrals, is taken into account in the prominence plasma.  We derive the dispersion relation of linear perturbations on the interface and analyze some limit cases analytically. Numerical results are obtained for realistic prominence parameters. We find that compressibility and gas pressure are important in determining the unstable flow velocities, specially in the range of sub-Alfv\'enic  flows that are consistent with the observations. The ambipolar diffusion has a generally destabilizing influence and reduces the threshold flow velocity for the KHi onset. 

\end{abstract}
\keywords{Instabilities; Magnetohydrodynamics; Prominences; Waves}
\end{frontmatter}

\section{Introduction}\label{s:introduction}

Solar prominences are one of the most interesting objects of the solar atmosphere. They consist of masses of relatively cool and dense plasma whose physical properties are similar to those in the solar chromosphere \citep[see, e.g.][]{vialengvold2015}. High-resolution observations have shown that their fine structure is composed by a myriad of thin ribbons called threads, which seem to outline particular  lines of the  magnetic structure \citep[see, e.g.][]{lin2011,martin2015prominences}. The dynamics of solar prominences is extremely complex and frequently observed phenomena include waves, flows, and instabilities \citep[see, e.g., the review by][]{arregui2018}.

%Observations have shown that waves are ubiquitous in fine structures of solar prominences. These waves are interpreted as magnetohydrodynamic (MHD) waves \citep[see][]{ballester2015waves}. According to the observations, there are evidences that waves are driven at the photosphere, where the feet of prominences magnetic field is anchored \citep{hillier2013}. It has been proved that waves can travel from the photosphere to coronal structures, such as solar prominences or coronal loops, transporting an important amount of energy. In homogeneous plasmas, MHD waves are classified into two different kinds: Alfv\'en waves, which are driven due to the forces produced by the magnetic field tension, and mangnetoacoustic waves, which are driven by the combined effect of  forces of the pressure gradient and the magnetic force. If the plasma is structured by the effect of the magnetic field, a new kind of MHD wave appears, which are called surface waves, which are defined by being confined to the vicinity of the interface where it propagates \citep[see the book by][]{roberts2019}.

The Kelvin-Helmholtz instability (KHi) is a classic fluid instability that may happen at the interface between two fluids in relative motion or in a continuous fluid with a shear flow. This instability is relevant in many astrophysical contexts, such as the magnetopause \citep{hasegawa2006}, planetary magnetospheres \citep{miura1984}, jets and outflows \citep{baty2006}, among others. Observations have shown that the KHi is also present in the solar atmosphere \citep[see, e.g.,][]{foullon2011,feng2013,kieokaew2021}. In solar prominences, the presence of KHi was first suggested by some works due to the observations of turbulent flows and vortices \citep[see, e.g.,][]{berger2010,ryutova2010}, and it was later confirmed by direct observations of its development \citep[see][]{hillier2018,yang2018}. 

The prominence core temperatures are estimated to be in the range 7000--9000~K, so that the plasma is partially ionized \citep[see][]{heinzel2015}. In a partially ionised plasma, charges and neutrals  have different stability properties with respect to the presence of a shear flow. Charges tend to be more stable owing to the role of magnetic tension, while neutrals are more prone to be unstable since they are not affected by the magnetic field. As both charges and neutrals are coupled through collisions, it is  found that a partially ionized plasma is generally more unstable to the presence of a shear flow than a fully ionized plasma.  A  review about the role of partial ionization on fluid instabilities, including the KHi, can be found in \citet{soler2022}.

The KHi onset in the presence of partial ionization has been modeled analytically in some previous works \citep[see, e.g.][]{watson2004,soler2012,martinez2015}. In these works, the two-fluid model was used, in which charges and neutrals are treated as separate fluids that interact through collisions. \cite{soler2012} performed a general study of the linear stage of the KHi at an interface between two partially ionized compressible plasmas in relative motion with a uniform magnetic field in the same direction as that of the fluid motion. In this situation, the classic linear analysis in fully ionized plasmas predicts an instability for super-Alfv\'enic flows alone. The results of \cite{soler2012} indicated that in partially ionized plasmas the KHi onset can happen for sub-Alfv\'enic flows owing to the presence of neutrals and their collisions with the charges. Subsequently, \cite{martinez2015} considered a cylindrical model aimed to represent a prominence thread with a longitudinal flow. They considered the incompressible approximation and found, again, instability for sub-Alfv\'enic flows. They concluded that collisions between charges and neutrals progressively reduce the KHi growth rate as the collision frequency increases, but the instability cannot be suppressed completely.

The high values of the charge-neutral collision frequency for prominence plasma conditions allows the use of the single-fluid MHD approximation in contrast to the more intricate two-fluid model \citep[see, e.g.,][]{Khomenko2014,ballester2018r}. In this context, the role of charge-neutral collisions is mainly present in the so-called ambipolar diffusion term. \citet{ballai2015} investigated incompressible surface waves on a prominence-corona interface with ambipolar diffusion in the prominence region and viscosity in the coronal region, although they considered an approximate treatment of the diffusion terms. They found instability for smaller flow velocities than those obtained in the absence of dissipation.  Recently, \citet{diaz2024} have used the single-fluid model to numerically explore  the KHi growth and turbulent evolution due to azimuthal flows caused by torsional oscillations in a cylindrical prominence thread model.

The aim of this work is to theoretically study the onset of the KHi in a  prominence-corona interface  with a shear flow. The effects of ambipolar diffusion, gas pressure, and compressibility are all considered together. Even though the single-fluid MHD model with ambipolar diffusion is simpler compared to the charged-neutral two-fluid model used previously by \citet{soler2012} and \citet{martinez2015}, it has never been analyzed fully in the context of the KHi onset. We perform a general mathematical derivation of the dispersion relation of linear perturbations in which the non-ideal terms are treated completely with no approximation. Limit cases of the dispersion relation are investigated analytically, while more generally applicable results are obtained numerically. The obtained results regarding the conditions for the KHi appearance are discussed. Specifically, a goal of this work is to determine whether the previous results obtained in the two-fluid case concerning the modification  of the instability threshold are dependent on the intricacies of two-fluid model or they can be recovered  with the  single-fluid MHD model with ambipolar diffusion.

\section{Model and basic equations}
\subsection{Interface model}
\begin{figure}
    \centering
    \includegraphics[width=0.8\textwidth]{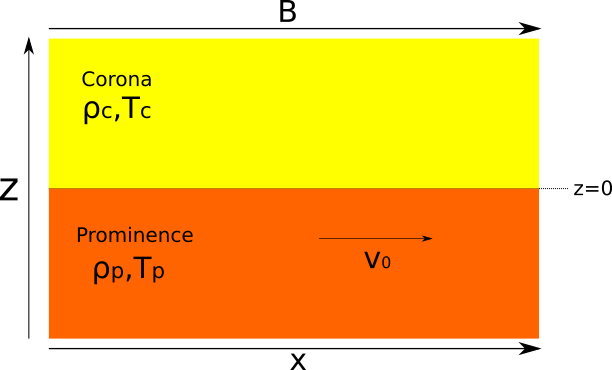}
    \caption{Sketch of the prominence-corona  interface model. All symbols are defined in the text.}
    \label{fig:model}
\end{figure}

We use a Cartesian interface model that aims to locally represent the boundary between a partially ionised solar prominence and the fully ionised solar corona. The  $z$-direction is normal to the interface and the plane $z=0$ corresponds to the interface itself, with the model being invariant in the $x$- and $y$-directions. The prominence and coronal regions are located at $z<0$ and $z>0$, respectively.   We consider a straight uniform magnetic field  aligned in the $x$-direction, namely $\vec{B}_{0} = B_{0} \hat{x}$, with a strength of $B_{0} =5$~G representative of quiescent prominences. In order to satisfy the requirement of  pressure balance across the interface, the gas pressure is taken to be uniform with a value of $p_{0}=5 \times 10^{-3}$~Pa.  A  sketch of the  model is shown in Figure \ref{fig:model}. Throughout the paper, the quantities related with the prominence region are denoted by the subscript $p$, while those related with the coronal region include the subscript $c$.

The equilibrium density profile depends on the $z$-direction through a piecewise uniform profile as
\begin{equation}
\rho_{0} =\left\{ \begin{array}{lll}
\rho_{\rm p}, & \textrm{if} & z \leq 0, \\
\rho_{\rm c}, & \textrm{if} & z > 0,
\end{array} \right.
\label{eq:density}
\end{equation}
meaning that there is an abrupt jump of the density in the interface.  The prominence density is computed with the ideal gas law, namely
\begin{equation}
    \rho_{\rm p} = \frac{m_{\rm p}}{k_{\rm B}} \frac{p_{0}}{T_{\rm p}} \tilde{\mu}, 
    \label{eq:gas}
\end{equation}
where $m_{\rm p}$ is the proton mass, $k_{\rm B}$ the Boltzmann constant, $T_{\rm p}=8000$~K is the assumed  temperature in the prominence region, $\tilde{\mu}$ is the mean atomic weight in the prominence, which for a hydrogen-helium plasma with a 10$\%$ of neutral helium and no helium ionization is given by
\begin{equation}
    \tilde{\mu} = \frac{1.4}{1.1+\xi_{\rm i}},
\end{equation}
where $\xi_{\rm i} = 0.52$ is the hydrogen ionisation fraction, i.e., the ratio of the electron number density to the total hydrogen number density, so that $\tilde{\mu} \approx 0.86$. The ionisation fraction is obtained from the tabulated values in \cite{heinzel2015} for an altitude of 20,000 km above the photosphere. The resulting prominence density is $\rho_{\rm p} \approx 6.5 \times 10^{-11}$~kg~m$^{-3}$. Concerning the density in the fully ionized coronal part, we used a density contrast of $\rho_{\rm p}/\rho_{\rm c}=100$, which is a typical density ratio between solar prominences and their surrounding corona. The resulting coronal temperature assuming full ionization of both hydrogen and helium (so that $\tilde{\mu} = 0.6$) is $T_{\rm c} \approx 5 \times 10^{5}$~K. 

It is important for our study to compute the characteristic velocities in the plasma. They are the Alfv\'en, $v_{\rm A}$, and sound, $c_{\rm s}$, velocities, given by
\begin{equation}
       v_{\rm A}=\frac{B_{0}}{\sqrt{\mu_{0}\rho_{0}}},\qquad
       c_{\rm s} = \sqrt{\frac{\gamma p_{0}}{\rho_{0}}},
\end{equation}
 where $\gamma$ is the adiabatic index and $\mu_{0}$ is the magnetic permeability. The values of the Alfv\'en velocities in the prominence and coronal regions are $v_{\rm Ap} \approx 55$~km~s$^{-1}$ and $v_{\rm Ac} \approx 550$~km~s$^{-1}$ respectively, while those of the sound velocities are $c_{\rm sp} \approx 11$~km~s$^{-1}$ and $c_{\rm sc} \approx 112$~km~s$^{-1}$.

We consider the effect of ambipolar diffusion in the partially ionised prominence region. The ambipolar diffusion coefficient, $ \eta_{\rm A}$, for a hydrogen-helium plasma is computed following \citet{zaqarashvili2013} and \citet{soler2015}, namely
\begin{equation}
    \eta_{\rm A} = \frac{\xi_{\rm n}^{2}\alpha_{\rm He}+\xi_{\rm He}^{2}\alpha_{\rm n}+2\xi_{\rm n}\xi_{\rm He}\alpha_{\rm n He}}{\mu_{0} \left( \alpha_{\rm He}\alpha_{\rm n}-\alpha_{\rm n He}^{2}\right)}, 
\end{equation}
where $\xi_{\rm n}=1-\xi_{\rm i} = 0.48$ is the hydrogen neutral fraction, $\xi_{\rm He}=0.4$ is the mass fraction of neutral helium, $\alpha_{\rm n}$ and $\alpha_{\rm He}$ are the neutral hydrogen and neutral helium total friction coefficients, respectively, and $\alpha_{\rm n He}$ is the symmetric friction coefficient for collisions between neutral hydrogen and neutral helium. The total friction coefficient of a species $\beta$ is computed as
\begin{equation}
    \alpha_{\rm \beta}= \sum_{\rm \beta \neq \beta'} \alpha_{\rm \beta \beta'},
\end{equation}
with $\alpha_{\rm \beta \beta'}$ the symmetric friction coefficient of collisions between species $\beta$ and $\beta'$, which could be neutral helium (He), neutral hydrogen (n), or protons (p). For collisions involving a neutral species $N$, so that $\beta' = N$, the symmetric friction coefficient can be computed as
\begin{equation}
    \alpha_{\rm \beta N} = n_{\rm \beta} n_{\rm N} m_{\rm \beta N} \sqrt{\frac{8 k_{\rm B}T_{\rm p}}{\pi m_{\rm \beta N}}} \sigma_{\rm \beta N},
\end{equation}
where $\sigma_{\rm \beta N}$ is the collisional cross-section and $m_{\rm \beta N}$ is the reduced mass. For the parameters considered, the ambipolar diffusion coefficient in the prominence plasma is  $ \eta_{\rm A} \approx 4 \times 10^{14}$~m$^{2}$~s$^{-1}$~T$^{-2}$. However, we note that the actual efficiency of the ambipolar diffusion depends upon the magnetic field strength \citep[see, e.g.,][]{Khomenko2012}, and the quantity that plays the role of a diffusion coefficient is $B_0^2 \eta_{\rm A}$. For the considered magnetic field strength of 5~G, we have $B_0^2 \eta_{\rm A} \approx 10^{8}$~m$^{2}$~s$^{-1}$.

Finally, we consider the presence of a shear flow at the interface. We assume that there is a field-aligned mass flow in the prominence region, while in the coronal region the plasma is static. Therefore, the mass flow velocity is
\begin{equation}
\vec{v}_{0} =\left\{ \begin{array}{lll}
v_{0}~\hat{x}, & \textrm{if} & z \leq 0, \\
0, & \textrm{if} & z > 0,
\end{array} \right.
\label{eq:flow}
\end{equation}
where $v_{0}$ denotes the flow velocity in the prominence, which is considered to be a constant. Observationally, the flow velocities observed in solar prominences are in the range 5--70~km~s$^{-1}$ \citep[see, e.g.][]{zirker1998,berger2010,mackay2010}. However,  we shall treat this flow velocity as a free parameter of the model.

\subsection{Governing equations}

We use the MHD equations for a partially ionized plasma in the single-fluid approximation. For the purpose of this investigation, we neglect all the non-ideal terms with the exception of the ambipolar diffusion. In order to study the onset of the instabilities driven by the shear flow, we linearise the equations by writing each variable as the sum of the background value and a small perturbation that depends on space and time. We shall use the subscript 0 to denote the background magnitude and the subscript 1 to denote the corresponding perturbation. The set of linearised equations are
\begin{equation}
        \rho_{0} \left(\frac{\partial \vec{v}_{1}}{\partial t} + \vec{v}_{0} \cdot \nabla \vec{v}_{1}\right) = - \nabla p_{1} + \frac{1}{\mu_{0}} \left( \nabla \times \vec{B}_{1} \right) \times \vec{B}_{0}, \label{eq:momentum}
\end{equation}
\begin{equation}
       \frac{\partial \vec{B}_{1}}{\partial t} = \nabla \times \left( \vec{v}_{0} \times \vec{B}_{1} \right) + \nabla \times \left( \vec{v}_{1} \times \vec{B}_{0} \right) + \eta_{\rm A}\nabla \times \{ \left( \nabla \times \vec{B}_{1} \right) \times \vec{B}_{0} \times \vec{B}_{0}\}, \label{eq:induction}
\end{equation}
 \begin{equation}
         \frac{\partial p_{1}}{\partial t} + \vec{v}_{0} \cdot \nabla p_{1} = -\gamma p_{0} \nabla \cdot v_{1}. \label{eq:energy}
 \end{equation}
 These equations are, respectively, the momentum equation, the induction equation and the energy equation.   For later use, it is also important to give the expression of the plasma Lagrangian displacement, $\vec{\xi}_{1}$, namely
\begin{equation}
    \frac{\partial \vec{\xi}_{1}}{\partial t} + \vec{v}_{0} \cdot \nabla \vec{\xi}_{1} = \vec{v}_{1}.
\end{equation}

Subsequently, we  express the perturbations as
\begin{equation}
    f_{1}(\vec{r},t) = f_{1}(z) \exp \left( st + i k_{x} + i k_{y}\right),
\end{equation}
where $s$ is the growth rate and $k_{x}$ and $k_{y}$ are the components of the wavenumber in the $x$- and $y$-directions, respectively. Essentially, we have performed a Fourier analysis in the $x$ and $y$ spatial directions and a Laplace analysis  in the temporal variable. We retain the full dependence of the perturbations in the $z$-direction due to the nonuniform nature  of the background.

We combine equations (\ref{eq:momentum})-(\ref{eq:energy}) and use  $\nabla \cdot \vec{v}_{1} $ as our main variable.  After some algebra, we arrive at the governing differential equation for the perturbations, namely
\begin{equation}
    \left( A \frac{\partial^{4}}{\partial z^{4}}-B\frac{\partial^{2}}{\partial z^{2}} + C\right)\nabla \cdot \vec{v}_{1} = 0,
    \label{eq:main}
\end{equation}
where $A$, $B$, and $C$ are coefficients given by
\begin{eqnarray}
    A &=& c_{\rm s}^{2} \left( \tilde{v}_{\rm A}^{2} - v_{\rm A}^{2} \right), \\
    B &=& (\tilde{v}_{\rm A}^{2}+c_{\rm s}^{2})\tilde{s}^{2}+2 k_{\perp}^{2}c_{\rm s}^{2} \left( \tilde{v}_{\rm A}^{2} - v_{\rm A}^{2} \right)+k_{x}^{2}c_{\rm s}^{2}v_{\rm A}^{2},\\
    C &=& \left( \tilde{s}^{2} + k_{x}^{2}c_{\rm s}^{2} \right)\left( \tilde{s}^{2} + k_{\perp}^{2}\tilde{v}_{\rm A}^{2} \right) + k_{y}^{2}c_{\rm s}^{2}\left( \tilde{s}^{2} + k_{\perp}^{2}\left( \tilde{v}_{\rm A}^{2} - v_{\rm A}^{2} \right) \right),
\end{eqnarray}
where $c_{\rm s}$ and $v_{\rm A}$ are the sound and Alfv\'en velocities defined before and $k_{\perp}^2 = k_{x}^{2}+k_{y}^{2}$ is the perpendicular wavenumber squared. Additionally, two important quantities that appear in these expressions are the Doppler-shifted growth rate,
\begin{equation}
    \tilde{s}=s + i k_{x}v_{0},
\end{equation}
and the square of the modified Alfv\'en velocity due to the presence of ambipolar diffusion \citep[see][]{forteza2008,soler2009},
\begin{equation}
    \tilde{v}_{\rm A}^{2} = v_{\rm A}^{2} + B_{0}^{2}\eta_{\rm A}\tilde{s}.
\end{equation}

In the ideal case, $\eta_{\rm A}=0$ and the modified Alfv\'en velocity, $\tilde{v}_{\rm A}$, reverts to the normal Alfv\'en velocity, $v_{\rm A}$. As a result of this, $A=0$ and the coefficients $B$ and $C$ adopt a more simplified form, namely 
\begin{eqnarray}
    B &=& (v_{\rm A}^{2}+c_{\rm s}^{2})\tilde{s}^{2}+k_{x}^{2}c_{\rm s}^{2}v_{\rm A}^{2},\\
    C &=& \left( \tilde{s}^{2} + k_{x}^{2}c_{\rm s}^{2} \right)\left( \tilde{s}^{2} + k_{\perp}^{2}v_{\rm A}^{2} \right) + k_{y}^{2}c_{\rm s}^{2}\tilde{s}^{2} ,
\end{eqnarray}
which allows us to rewrite Equation~(\ref{eq:main}) as
\begin{equation}
    \left( \frac{\partial^{2}}{\partial z^{2}} - \left( \mathbb{K}^{2} + k_{y}^{2} \right) \right) \nabla \cdot \vec{v}_{1} = 0,
    \label{eq:ideal}
\end{equation}
where now the quantity
\begin{equation}
    \mathbb{K}^{2}  =\frac{\left( \tilde{s}^{2} + k_{x}^{2}c_{\rm s}^{2} \right)\left( \tilde{s}^{2} + k_{x}^{2}v_{\rm A}^{2} \right)}{(v_{\rm A}^{2}+c_{\rm s}^{2})\tilde{s}^{2}+k_{x}^{2}c_{\rm s}^{2}v_{\rm A}^{2}},
\end{equation}
can be related to an effective wavenumber in the $z$-direction. The ideal Equation~({\ref{eq:ideal}})  was already obtained in previous works \citep[see, e.g.,][]{roberts1981,jain1991}. However, the governing equation that includes ambipolar diffusion (Equation~(\ref{eq:main})) has not been derived before in the literature.

\subsection{Dispersion relation}

The goal is to obtain the dispersion relation for the linear perturbations at the interface. To do so, we seek  solutions in the form of evanescent perturbations that decay away from the interface. 

Equation~(\ref{eq:main}) applies in the prominence region ($z<0$) where ambipolar diffusion is present. Substituting a solution of the form $\nabla \cdot \vec{v}_{1} \sim \exp\left( k_{\rm p} z\right)$ in Equation~(\ref{eq:main}) results in two possible values of $k_{\rm p}$, namely $k_{\rm p+}$ and $k_{\rm p-}$, given by
\begin{equation}
    k_{\rm p\pm}^{2} = k_{y}^{2} + \frac{1}{2}\frac{\left( \tilde{v}_{\rm A}^{2} + c_{\rm s}^{2} \right)\tilde{s}^{2}+k_{x}^{2}c_{\rm s}^{2}\left( 2\tilde{v}_{\rm A}^{2} - v_{\rm A}^{2} \right)}{c_{\rm s}^{2}\left( \tilde{v}_{\rm A}^{2} - v_{\rm A}^{2} \right)}\left( 1 \pm \sqrt{1-\delta}\right), \label{eq:kp}
\end{equation}
where
\begin{equation}
    \delta = 4c_{\rm s}^{2}\left( \tilde{v}_{\rm A}^{2} - v_{\rm A}^{2} \right)\frac{\left( \tilde{s}^{2} + k_{x}^{2}c_{\rm s}^{2} \right)\left( \tilde{s}^{2} + k_{x}^{2}\tilde{v}_{\rm A}^{2} \right)}{\left(\left( \tilde{v}_{\rm A}^{2} + c_{\rm s}^{2} \right)\tilde{s}^{2}+k_{x}^{2}c_{\rm s}^{2}\left( 2\tilde{v}_{\rm A}^{2} - v_{\rm A}^{2} \right)\right)^{2}}. \label{eq:delta}
\end{equation}
\\
In Equations~(\ref{eq:kp}) and (\ref{eq:delta}) all quantities should be computed using prominence conditions. Conversely, in the dissipationless, ideal coronal region ($z >0$) it is enough to consider Equation~({\ref{eq:ideal}}). From substituting again a solution of the form $\nabla \cdot \vec{v}_{1} \sim \exp\left(- k_{\rm c} z\right)$ in Equation~(\ref{eq:ideal}), we obtain 
\begin{equation}
    k_{\rm c}^2 =\mathbb{K}^{2}+k_{y}^{2} = \frac{\left( \tilde{s}^{2} + k_{x}^{2}c_{\rm s}^{2} \right)\left( \tilde{s}^{2} + k_{x}^{2}v_{\rm A}^{2} \right)}{(v_{\rm A}^{2}+c_{\rm s}^{2})\tilde{s}^{2}+k_{x}^{2}c_{\rm s}^{2}v_{\rm A}^{2}} + k_y^2,\label{eq:kc}
\end{equation}
where all quantities should be computed using coronal conditions.

Therefore, the general solution for $\nabla \cdot \vec{v}_{1}$ in the two regions takes the form of
\begin{equation}
\nabla \cdot \vec{v}_{1} =\left\{ \begin{array}{lll}
A_{1}\exp \left( k_{\rm p+}z \right)+A_{2}\exp \left( k_{\rm p-}z \right), & \textrm{if} & z \leq 0, \\
A_3\exp \left( -k_{\rm c} z \right), & \textrm{if} & z > 0,
\end{array} \right.
\label{eq:compressibility}
\end{equation}
where $A_{1}$, $A_{2}$ and $A_3$ are constants, $k_{p \pm}$ represent the effective wavenumbers in the $z$-direction in the prominence region, which are computed from Equation~(\ref{eq:kp}), while $k_{\rm c}$ is the $z$-directed effective wavenumber in the coronal region, obtained from Equation~(\ref{eq:kc}). In order to obtain the dispersion relation, we need to impose  boundary conditions on the interface. These conditions are the continuities of the normal component of the Lagrangian displacement, $\xi_{1z} = v_{1z}/\tilde{s}$, the total pressure perturbation, $p'_{T} = p_{1} + B_{0}B_{1x}/\mu_{0}$, and the normal component of the magnetic field perturbation, $B_{1z}$. The continuity of the normal component of the Lagrangian displacement is justified by the need that the transverse motions to the interface maintain the fluid coherence, while the continuity of the total pressure perturbation is based on the need of a force balance \citep[see, e.g.,][]{roberts2024}. The presence of the ambipolar diffusion requires  an additional boundary condition, which is the electrodynamic condition that $B_{1z}$ must be continuous \citep[see, e.g.,][]{tomimura1990}. The continuity of $B_{1z}$ is automatically satisfied in the ideal case because of the continuity of $\xi_{1z}$. However, in presence of the ambipolar diffusion, the continuity of $\xi_{1z}$ does not result in the continuity of $B_{1z}$ due to additional terms proportional to the ambipolar diffusion coefficient. Then, we have to explicitly enforce  this condition.

By combining Equations~(\ref{eq:momentum})-(\ref{eq:energy}), we can express $\xi_{1z}$, $p'_{T}$ and $B_{1z}$ as functions of $\nabla \cdot \vec{v}_{1}$. In the dissipative prominence region, they are expressed as
\begin{eqnarray}
    \xi_{\rm 1z} &=& \frac{ \left[\left( c_{\rm s}^{2} + \tilde{v}_{\rm A}^{2} \right)\tilde{s}^{2}+k_{\perp}^{2}c_{\rm s}^{2}\left( \tilde{v}_{\rm A}^{2}-v_{\rm A}^{2} \right)+k_{x}^{2}c_{\rm s}^{2}\tilde{v}_{\rm A}^{2}\right]\frac{\partial}{\partial z}-c_{\rm s}^{2}\left( \tilde{v}_{\rm A}^{2}-v_{\rm A}^{2} \right)\frac{\partial^{3}}{\partial z^{3}}}{\tilde{s}^{3}\left( \tilde{s}^{2}+k_{x}^{2}\tilde{v}_{\rm A}^{2} \right)}\nabla \cdot \vec{v}_{1},
    \nonumber \\ \label{eq:ldp} \\
    p'_{\rm T} &=& -\frac{\rho_{0}}{\tilde{s}^{3}}\left( \left[ \left(c_{\rm s}^{2}+\tilde{v}_{\rm A}^{2}\right)\tilde{s}^{2}+k_{\perp}^{2}c_{\rm s}^{2}\tilde{v}_{\rm A}^{2} -k_{y}^{2}c_{\rm s}^{2}v_{\rm A}^{2}\right]-c_{\rm s}^{2}\left(\tilde{v}_{\rm A}^{2}-v_{\rm A}^{2}\right)\frac{\partial^{2}}{\partial z^{2}} \right)\nabla \cdot \vec{v}_{1},
    \label{eq:ptp} \\
        B_{\rm 1z} &=& \frac{i k_{x} B_{0}}{\tilde{s}^{2}+k_{x}^{2}\tilde{v}_{\rm A}^{2}}\Bigl( \left[ \frac{c_{\rm s}^{2}}{\tilde{s}} + \frac{\tilde{v}_{\rm A}^{2}}{\tilde{s}^{3}v_{\rm A}^{2}}\left( k_{y}^{2}c_{\rm s}^{2}\left( \tilde{v}_{\rm A}^{2} - v_{\rm A}^{2} \right) +\left(\tilde{s}^{2}+k_{x}^{2}c_{\rm s}^{2}\right) \tilde{v}_{\rm A}^{2}  \right)\right] \frac{\partial}{\partial z} \nonumber \\
        &&-\frac{\tilde{v}_{\rm A}^{2}}{\tilde{s}^{3}}\frac{c_{\rm s}^{2}\left( \tilde{v}_{\rm A}^{2} - v_{\rm A}^{2} \right)}{v_{\rm A}^{2}}\frac{\partial^{3}}{\partial z^{3}} \Bigr)  \nabla \cdot \vec{v}_{1}.
        \label{eq:bzp}
\end{eqnarray}
In the ideal coronal region, the absence of  ambipolar diffusion reduces the Equations (\ref{eq:ldp})-(\ref{eq:bzp}) to
\begin{eqnarray}
    \xi_{\rm 1z} &=& \frac{\left( c_{\rm s}^{2} + v_{\rm A}^{2} \right)\tilde{s}^{2}+k_{x}^{2}c_{\rm s}^{2}v_{\rm A}^{2}}{\tilde{s}^{3}\left( \tilde{s}^{2}+k_{x}^{2}v_{\rm A}^{2} \right)}\frac{\partial}{\partial z}\nabla \cdot \vec{v}_{1},
    \label{eq:ldc} \\
    p'_{\rm T} &=& -\frac{\rho_{0}}{\tilde{s}^{3}}\left( \left( c_{\rm s}^{2} + v_{\rm A}^{2} \right)\tilde{s}^{2} +k_{x}^{2} c_{\rm s}^{2}v_{\rm A}^{2} \right)\nabla \cdot \vec{v}_{1},
    \label{eq:ptc} \\
    B_{\rm 1z} &=& ik_{x}B_{0}\frac{\left( c_{\rm s}^{2} + v_{\rm A}^{2} \right)\tilde{s}^{2}+k_{x}^{2}c_{\rm s}^{2}v_{\rm A}^{2}}{\tilde{s}^{3}\left( \tilde{s}^{2}+k_{x}^{2}v_{\rm A}^{2} \right)}\frac{\partial}{\partial z}\nabla \cdot \vec{v}_{1} = ik_{x}B_{0}\xi_{1z}.
    \label{eq:bzc}
\end{eqnarray}
The presence of  ambipolar diffusion results in some additional terms when the expressions of $\xi_{1z}$, $p'_{T}$ and $B_{1z}$ are compared with those in the ideal case. Those terms involve derivatives of higher order in $z$. Importantly, we note again that in the ideal case, the continuity of $B_{1z}$ is already satisfied by the continuity of $\xi_{1z}$, while in the ambipolar case we must enforce this additional condition.

By imposing  the continutiy conditions at the interface, a system of three equations for three unknowns, namely $A_{1}$, $A_{2}$ and $A_3$, is obtained. The non-trivial solution of this system provides us with the dispersion relation,
\begin{equation}
      \mathcal{D}(s) = 0,
\end{equation}
where $\mathcal{D}(s)$ is the dispersion function. The procedure to arrive at the full expression of $\mathcal{D}(s)$ is included in the Appendix~\ref{app}.

\section{Results}

Before analysing the solutions of the full dispersion relation, we first discuss some limit cases that can help us understand the physics of the results. These limit cases are the incompressible limit and the compressible pressureless case. The aim of this approach is to progressively increase the degree of complexity,  in order to understand better the specific role of each effect.

\subsection{Incompressible case}

The  incompressible limit can be achieved by taking $c_{\rm s} \rightarrow \infty$ in both regions. In this limit, the effective wavenumbers in the prominence and coronal regions reduce to,
\begin{eqnarray}
        k_{\rm p+}^{2} &=& k_{y}^{2}+\frac{\tilde{s}^{2}+k_{x}^{2}\tilde{v}_{\rm A}^{2}}{\tilde{v}_{\rm A}^{2}-v_{\rm A}^{2}}, \\
    k_{\rm p-}^{2} &=& k_{y}^{2}+k_{x}^{2} = k_{\perp}^{2}, \\
    k_{\rm c}^{2} &=& k_{y}^{2}+k_{x}^{2} = k_{\perp}^{2},
\end{eqnarray}
and the dispersion relation simplifies to,
\begin{equation}
    \mathcal{D}(s) = \rho_{\rm p}(\tilde{s}^{2}+k_{x}^{2}v_{\rm Ap}^{2}) + \rho_{\rm c}(s^{2}+k_{x}^{2}v_{\rm Ac}^{2})=0.
    \label{eq:incompressible}
\end{equation}
This dispersion relation is the same  as that for incompressible ideal surface waves  \citep[see, e.g.,][]{roberts1981,nakariakov1995}. The incompressible approximation suppresses the  compressive magnetosonic modes, and therefore the dispersion relation  has a reduced order compared to that in the complete case. An important result is the absence in Equation~(\ref{eq:incompressible}) of any term related with the ambipolar diffusion. This fact anticipates that, in the incompressible limit, the onset of shear-flow instabilities is not affected by the ambipolar diffusion.

In absence of mass flow, $\tilde{s} = s$ and the solution of Equation~(\ref{eq:incompressible}) is
\begin{equation}
    s = \pm i k_{x} \sqrt{\frac{\rho_{\rm p}v_{\rm Ap}^{2}+\rho_{\rm c}v_{\rm Ac}^{2}}{\rho_{\rm p}+\rho_{\rm c}}} \equiv \pm i\, \omega_{\rm k}. \label{eq:wk}
\end{equation}
The growth rate is purely imaginary, which means that this solution is oscillatory in time with frequency $\omega_{\rm k}$. In fact, this is the well known frequency of incompressible surface waves: the kink frequency \citep[see, e.g.][]{roberts1981,edwin1982}. The $\pm$ sign in Equation~(\ref{eq:wk}) corresponds to forward/backward surface waves with respect to the direction of the magnetic field, respectively. No instability of these surface waves can happen in the absence of flow.

In contrast, if the flow is present in the prominence region, the solution of Equation~(\ref{eq:incompressible}) is
\begin{equation}
  s = -i k_x\frac{\rho_{\rm p}v_{0}}{\rho_{\rm p}+\rho_{\rm c}} \pm i  \sqrt{\omega_{\rm k}^2-\frac{\rho_{\rm p}\rho_{\rm c}}{(\rho_{\rm p}+\rho_{\rm c})^{2}}k_{x}^2v_{0}^{2}}.
    \label{eq:freqinc}
\end{equation}
Comparing Equations~(\ref{eq:wk}) and (\ref{eq:freqinc}), we note now  the presence of  a purely imaginary advection term proportional to $v_0$ (the first term on the right-hand side of Equation~(\ref{eq:freqinc})), which produces a Doppler shift to the oscillation frequency. Besides this advection term, there is a negative correction term within the square root in Equation~(\ref{eq:freqinc}). Indeed, the sign of the square root argument informs us about the stability of the perturbations. The new negative term overpowers the originally present positive term when $v_0 > v_{\rm KH}$, with
\begin{equation}
    v_{\rm KH} = \sqrt{\frac{\rho_{\rm p}+\rho_{\rm c}}{\rho_{\rm p}\rho_{\rm c}}(\rho_{\rm p}v_{\rm Ap}^{2}+\rho_{\rm c}v_{\rm Ac}^{2})} = \sqrt{\frac{2 B_0^2}{\mu_0}\frac{\rho_{\rm p}+\rho_{\rm c}}{\rho_{\rm p}\rho_{\rm c}}} = \sqrt{\frac{2(\rho_{\rm p}+\rho_{\rm c})}{\rho_{\rm c}}}\, v_{\rm Ap}.
    \label{eq:vcritin}
\end{equation}
Then, the growth rate acquires a real part. Because of the $\pm$ sign in front of the square root,  there is a unstable solution with ${\rm Re}(s) > 0$ and another stable solution with ${\rm Re}(s) < 0$. The unstable solution is the classic KHi and $v_{\rm KH}$ is the flow velocity threshold for the KHi onset \citep[see][]{chandrasekhar1961,soler2012,martinez2015}. We note that this critical velocity is independent of $k_x$.

\begin{figure}
    \centering
    \includegraphics[width=0.8\linewidth]{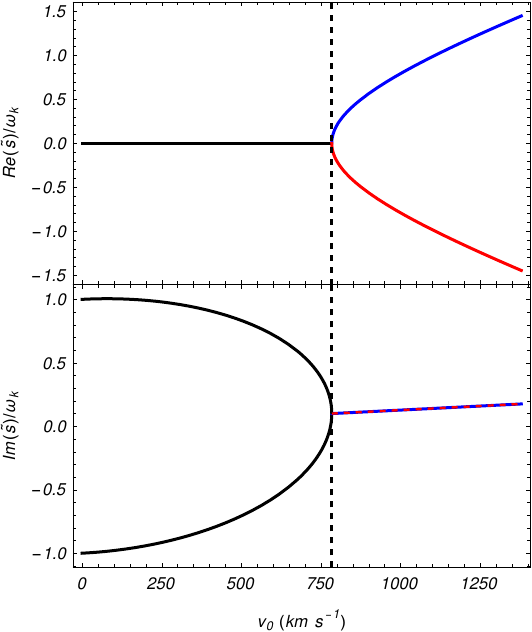}
    \caption{Real part (top) and imaginary part (bottom) of the normalized Doppler-shifted growth rate as a function of the  flow velocity in the incompressible case. The vertical  dashed line denotes the KHi critical velocity of Equation~(\ref{eq:vcritin}). The blue and red lines after the critical velocity correspond to the unstable and stable branches, respectively.}
    \label{fig:inc}
\end{figure}

To illustrate the transition to the unstable regime, the real and imaginary parts  of  Doppler-shifted growth rate obtained from Equation~(\ref{eq:freqinc}) are plotted in Figure~\ref{fig:inc} as functions of the flow velocity.  We plot the Doppler-shifted growth rate, $\tilde{s}$, instead of the actual growth rate, $s$, to partially remove the advection effect due to the flow. This advection modifies the imaginary part, but not the real part of the growth rate. The critical flow of Equation~(\ref{eq:vcritin}) is indicated with a vertical line in Figure~\ref{fig:inc}. The real part of the Doppler-shifted growth rate, plotted in the upper panel of Figure \ref{fig:inc}, is zero for $v_0 < v_{\rm KH}$, showing the absence of either instability or damping for slow flows. Conversely, for flow velocities above the critical one, two distinct branches appear, with the blue line representing the unstable branch and the red line the stable (damped) branch. On the other hand, the imaginary part of the Doppler-shifted growth rate is represented in the lower panel of Figure~\ref{fig:inc}. The two solutions present for flows below the critical velocity correspond to the forward-propagating and backward-propagating surface waves. From the critical flow onwards, the two waves converge and the KHi appears.

For the physical conditions in the model, the critical velocity according to Equation~(\ref{eq:vcritin}) is $v_{\rm KH} \approx 784$~km~s$^{-1}$, which is larger than the typically observed flows in prominences. As the KHi has been observed in prominences \citep[see, e.g.][]{zirker1998,berger2010}, the applicability of the incompressible limit to this context can be questioned. The absence of any effect related to ambipolar diffusion is also an argument against the validity of the incompressible approximation. In addition to that, the present incompressible results in the single-fluid approximation seem to be in apparent contradiction with the previous findings of \cite{martinez2015}. They also considered the incompressible limit, but obtained an instability for slower flows than $v_{\rm KH}$. The reason for this different result is that  \cite{martinez2015} used the two-fluid theory, in which neutrals are treated as a separate fluid from the ionized fluid (see details in the review by \citet{soler2024}). The growth rate of the instability for slow flows found by \cite{martinez2015} is inversely proportional to the collision frequency between neutrals and ions (see their Equation~(30)). The single-fluid approximation used here is equivalent to taking very large values of the collision frequency, which would result in a vanishing growth rate according to the expression of \cite{martinez2015}. This means that the present incompressible results do not contradict but agree with those of \cite{martinez2015} when the appropriate limit is considered in the results of \cite{martinez2015}.

\subsection{Compressible pressureless case}

Another interesting limit to investigate is the pressureless case, in which the plasma compressibility is taken into account but the thermal pressure is neglected. The pressureless case is an intermediate step between the incompressible limit and the complete case, in the sense that the role of compressibility is present but acoustic effects remain absent. Investigating this case allows us to determine the effect of compressibility alone. By setting the background pressure as $p_{0}=0$, the sound speed vanishes: $c_{\rm s}=0$. As a consequence of that, the effective wavenumbers in the prominence and coronal regions become,
\begin{eqnarray}
     k_{\rm p+}^2 & \to & \infty, \\
    k_{\rm p-}^2 &=& k_{y}^{2}+\frac{\tilde{s}^{2}+k_{x}^{2}\tilde{v}_{\rm Ap}^{2}}{\tilde{v}_{\rm Ap}^{2}},\\
    k_{\rm c}^2 &=& k_{y}^{2}+\frac{s^{2}+k_{x}^{2}v_{\rm Ac}^{2}}{v_{\rm Ac}^{2}}.
\end{eqnarray}
 The dispersion relation in this limit is
\begin{equation}
        \mathcal{D}(s) = \rho_{\rm p}(\tilde{s}^{2}+k_{x}^{2}\tilde{v}_{\rm Ap}^{2})k_{\rm c} + \rho_{\rm c}(s^{2}+k_{x}^{2}v_{\rm Ac}^{2})k_{\rm p-}=0.
        \label{eq:beta0}
\end{equation}
Contrary to the dispersion relation in the  incompressible limit (Equation~(\ref{eq:incompressible})), now there are terms in Equation~(\ref{eq:beta0}) that include the effect of ambipolar diffusion. These dissipative terms are enclosed in the definitions of $\tilde{v}_{\rm Ap}$ and $k_{\rm p-}$. It is also important to note that, in the presence of compressibility, the surface wave solutions to Equation~(\ref{eq:beta0}) are fast magnetoacoustic waves. Slow magnetoacoustic waves are absent owing to the vanishing sound speed.

Some analytic progress can be made in the case of nearly perpendicular propagation, so that $k_y \gg k_x$, and in the absence of flow, $v_0=0$. In this particular scenario, the solution of Equation~(\ref{eq:beta0}) is
\begin{equation}
    s =-\frac{\rho_{\rm p}B_{0}^{2}\eta_{\rm A}}{2(\rho_{\rm p}+\rho_{\rm c})}k_{x}^{2} \pm i \sqrt{\omega_{\rm k}^2-\frac{\rho_{\rm p}^{2}B_{0}^{4}\eta_{\rm A}^{2}}{4(\rho_{\rm p}+\rho_{\rm c})^{2}}k_{x}^{4}}.
    \label{eq:freqbeta}
\end{equation}
The growth rate is complex, with a negative real part proportional to the ambipolar coefficient. This accounts for the damping of the perturbations associated with the ambipolar diffusion. As before, the imaginary part of the growth rate corresponds to the frequency of the surface waves, which is now modified by the ambipolar diffusion. For a particular value of $k_x$, namely 
\begin{equation}
    k_{x}^{\rm cut.} = \frac{2\sqrt{\rho_{\rm p}+\rho_{\rm c}}\sqrt{\rho_{\rm p}v_{\rm Ap}^{2}+\rho_{\rm c}v_{\rm Ac}^{2}}}{\rho_{\rm p}B_{0}^{2}\eta_{\rm A}},
    \label{eq:cutoff}
\end{equation}
the imaginary part of the growth rate vanishes and  the growth rate becomes real for $k_x > k_{x}^{\rm cut.}$. However, there is no instability in this case as both $+$ and $-$ signs in Equation~(\ref{eq:freqbeta}) always give a negative real part of $s$, i.e., a damped disturbace. This particular value of $k_x$ is the so-called cut-off wavenumber for which ambipolar diffusion produces the critical damping of the waves. A detailed comment on the nature of this cut-off wavenumber is given in \citet{soler2024}. For the considered model parameters,  $k_{x}^{\rm cut.} \approx 1.5 \times 10^{-3}$ m$^{-1}$, which corresponds to a wavelength of $\approx 4.2$~km. Such a small wavelength is well below the resolution capability of current instruments, so that the existence of this cut-off wavenumber is of no practical importance for this investigation.

Next, we incorporate the effect of the flow, so that $v_0\neq 0$, but still retain the limit of nearly perpendicular propagation, $k_y \gg k_x$. In this case, the solution of Equation~(\ref{eq:beta0}) is
\begin{eqnarray}
        s &=&-\frac{\rho_{\rm p}B_{0}^{2}\eta_{\rm A}}{2(\rho_{\rm p}+\rho_{\rm c})}k_{x}^{2} - ik_x\frac{ \rho_{\rm p}v_{0}}{(\rho_{\rm p}+\rho_{\rm c})} \nonumber \\ 
        && \pm i\sqrt{\omega_{\rm k}^{2}-\frac{4\rho_{\rm p}\rho_{\rm c}k_x^2 v_0^2+\rho_{\rm p}^2 k_x^4 B_{0}^{4}\eta_{\rm A}^2}{4(\rho_{\rm p}+\rho_{\rm c})^2}+i\frac{\rho_{\rm p}\rho_{\rm c}k_x^3 B_{0}^{2}\eta_{\rm A}v_0}{(\rho_{\rm p}+\rho_{\rm c})^2}}.
        \label{eq:betafreq}
\end{eqnarray}
The  terms on the right-hand side of Equation~(\ref{eq:betafreq}) are already familiar to us. The first two terms  correspond to the ambipolar damping and the flow Doppler shift, respectively. The third term (that with the square root) is the surface wave frequency modified by both ambipolar diffusion and flow. In fact, we see that the joint acction of the two effects gives rise to an imaginary contribution within the square root, which complicates the mathematical analysis of the expression.  In the case that the ambipolar diffusion is dropped, $\eta_{\rm A} = 0$, the critical flow velocity is exactly the same as that found in the incompressible limit (Equation~(\ref{eq:vcritin})). In the presence of ambipolar diffusion, $\eta_{\rm A} \neq 0$, the  calculation is more involved, but  it is still possible to obtain an expression of the critical flow velocity for which the real part of the growth rate becomes positive, denoting an instability. The procedure consists in finding the flow velocity for which the positive imaginary part of the square root term becomes larger than the negative imaginary part due to the ambipolar damping term. After some algebraic manipulations, the critical velocity in the ambipolar case turns out to be,
\begin{equation}
    v_{\rm amb.} = \sqrt{\frac{\rho_{\rm p}v_{\rm Ap}^{2}+\rho_{\rm c}v_{\rm Ac}^{2}}{\rho_{\rm c}}} = \sqrt{\frac{2 B_0^2}{\mu_0 \rho_{\rm c}}}=\sqrt{\frac{2\rho_{\rm p}}{\rho_{\rm c}}}\, v_{\rm Ap}.
    \label{eq:vcritb}
\end{equation}
The relation between the critical velocity in the ideal case (Equation~(\ref{eq:vcritin})) and that in the ambipolar case (Equation~(\ref{eq:vcritb})) is
\begin{equation}
    v_{\rm amb.} = \sqrt{\frac{\rho_{\rm p}}{\rho_{\rm p}+\rho_{\rm c}}}\,v_{\rm KH}, \label{eq:relvels}
\end{equation}
so that $v_{\rm amb.} < v_{\rm KH}$. For the considered density contrast in the model, $v_{\rm amb.} \approx 0.995\, v_{\rm KH} \approx 780$~km~s$^{-1}$.

Now, we can draw some observations about the instability onset for nearly perpendicular propagation. In the absence of ambipolar diffusion, this case turns out to be equivalent to the incompressible limit whose critical flow velocity is in Equation~(\ref{eq:vcritin}). Conversely, when ambipolar diffusion is present, the incompressible critical flow velocity of Equation~(\ref{eq:vcritin}) is replaced with that  of Equation~(\ref{eq:vcritb}), which is slightly smaller. Although the difference in the critical velocity is rather negligible, one may conclude that the effect of ambipolar diffusion is to reduce the critical flow velocity. Nevertheless,  Equation~(\ref{eq:vcritb}) is  independent of the value of the ambipolar coefficient itself. As long as ambipolar diffusion is present, however small the ambipolar coefficient is, the critical flow velocity is that of Equation~(\ref{eq:vcritb}). But if ambipolar diffusion is dropped, the critical flow velocity immediately reverts to that of Equation~(\ref{eq:vcritin}). 

% \begin{figure}
%     \centering
%   \includegraphics[width=0.8\textwidth]{vamb.eps}
%    \caption{Critical mass flow as function of the ratio between the coronal amb prominence ambipolar diffusion coefficient. The dashed line represents the result of Equation \ref{eq:vcritin}, while the dashed-dotted line represents the result of Equation \ref{eq:vcritb}.}
%    \label{fig:vamb}
% \end{figure}

This discontinuous behavior of  the critical flow velocity was already found in previous studies where the considered dissipation mechanism was not ambipolar diffusion but viscosity \citep[e.g.,][]{ruderman1995,ruderman1996}. However, in our view, the discontinuous behavior of  the critical flow velocity as a function of the ambipolar coefficient at $\eta_{\rm A} = 0$ is inconsistent from the physical perspective, as  a continuous  behavior of the critical flow when $\eta_{\rm A} \to 0$ should be expected. The source of the inconsistency in the present and previous works resides in the model set-up, where an abrupt boundary is considered with dissipation acting on one side of the boundary alone. Replacing the abrupt boundary by a smooth boundary or  considering dissipation on both sides of the abrupt boundary would resolve the inconsistency. To illustrate this, we have repeated the  calculation of the critical velocity but   including a nonzero ambipolar coefficient  in the coronal part. In reality, diffusion of neutrals across the boundary would produce a small but nonzero ambipolar coefficient in the coronal plasma adjacent to the boundary. The critical flow velocity in this case turns out to be, 
\begin{equation}
    v_{\rm amb.} = v_{\rm KH} \left[ 1 + \rho_{\rm p}\rho_{\rm c}\left(\frac{\eta_{\rm A} - \eta_{\rm A,c}}{\rho_{\rm p}\eta_{\rm A} + \rho_{\rm c}\eta_{\rm A,c}} \right)^2 \right]^{-1/2}, \label{eq:critgen}
\end{equation}
where $\eta_{\rm A,c}$ now denotes the ambipolar coefficient in the  coronal region. In the case that $\eta_{\rm A,c} = 0$, the dependence on the prominence plasma $\eta_{\rm A}$ cancels out and Equation~(\ref{eq:critgen}) reverts to Equation~(\ref{eq:relvels}). Since  $\rho_{\rm p} \gg \rho_{\rm c}$, Equation~(\ref{eq:critgen}) can approximately be written as,
\begin{equation}
    v_{\rm amb.} \approx  v_{\rm KH} \left[ 1 + \frac{\rho_{\rm c}}{\rho_{\rm p}} \left(1 - \frac{\eta_{\rm A,c}}{\eta_{\rm A}} \right)^2 \right]^{-1/2}, \label{eq:critgen2}
\end{equation}
which more explicitly shows the dependence of the critical flow on the ambipolar coefficient. To consistently recover the ideal threshold, one sees that the ratio $\eta_{\rm A,c} / \eta_{\rm A}$ should naturally tend to 1 in the limit of vanishing ambipolar diffusion. 

%Figure \ref{fig:vamb} shows the variation of the critical flow velocity from the case in which there is only ambipolar diffusion in the prominence region to the one in which the ambipolar coefficient is the same in both regions.

\begin{figure}
    \centering
  \includegraphics[width=0.8\textwidth]{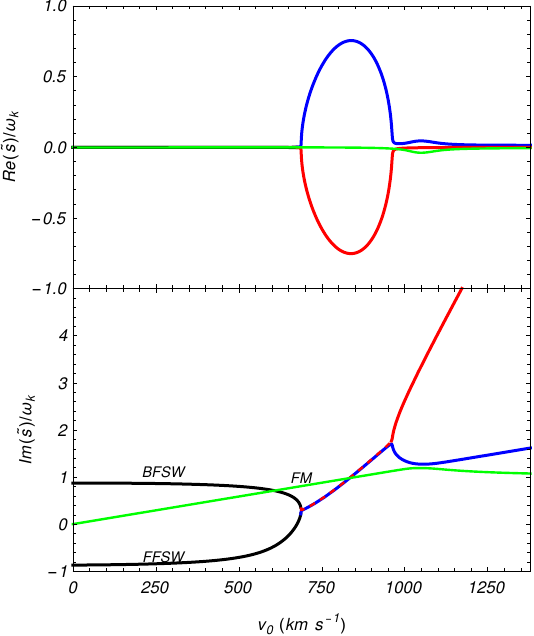}
   \caption{Real part (top) and imaginary part (bottom) of the normalized Doppler-shifted growth rate as a function of the  flow velocity in the compressible pressureless case for $k_{x} =k_{y}=\pi/L$, with $L =$~1,000~km.}
   \label{fig:betakh}
\end{figure}

In connection to this effect caused by dissipation, some studies have attributed  the presence of instability for flow velocities below the ideal threshold, $v_{\rm KH}$, but above the dissipative threshold, $v_{\rm amb.}$ in our case, to a so-called dissipative instability that would be unrelated to the KHi. We disagree with this interpretation. In our calculations,  dissipation does not cause a new instability, but it merely decreases the velocity threshold for the KHi onset, as the above study evidences. 

The stability analysis beyond the limit of nearly perpendicular propagation requires the numerical solution of the dispersion relation (Equation~(\ref{eq:beta0})). The investigation is more involved, as departing from the limit $k_y \gg k_x$ introduces the effect of compressibility \citep[see][and references therein]{soler2012}.  Figure~\ref{fig:betakh} shows the result of the numerical solution of Equation~(\ref{eq:beta0}) as a function of the flow velocity for $k_y/k_x = 1$. While in the ideal case the particular values of the wavenumber components are irrelevant for the stability (only their ratio is important), in the presence of ambipolar diffusion the actual value of $k_x$ determines the efficiency of the dissipation. In these calculations, we have considered $k_x = \pi/L$, where $L$ is a  length scale along the magnetic field lines. In the context of solar prominences, this length scale can be related to a characteristic length of the prominence threads \citep[see][]{terradas2021,melis2023}. Consequently, we set $L =$~1,000~km. This gives a value of $k_x$ that is about three orders of magnitude smaller than $k_{x}^{\rm cut.}$.

%This allows the dispersion relation to be satisfied for a new particular solution that does not exist when the flow is absent. 

Besides the forward and backward fast surface waves, labeled by FFSW and BFSW in Figure~\ref{fig:betakh}, Equation~(\ref{eq:beta0}) admits an additional solution that was trivial in the incompressible limit and now verifies that $k_{\rm p-} =- k_{\rm c}$. In the ideal case, the growth rate of this mode is purely imaginary, namely
\begin{equation}
    s = -i k_{x} \frac{v_{\rm Ac}}{v_{\rm Ac}+v_{\rm Ap}} v_{0}.
    \label{eq:doppler}
\end{equation}
In the presence of ambipolar diffusion, this mode acquires a small negative real part of the growth rate, but this is not relevant for our discussion. For simplicity, we call this solution the flow mode, FM, and is also plotted in Figure~\ref{fig:betakh} with a green line. The FM is originally a non-propagating mode that is advected by the flow. Therefore, it appears as a forward-propagating solution, but there is no backward-propagating counterpart.

\begin{figure}
    \centering
    \includegraphics[width=0.8\textwidth]{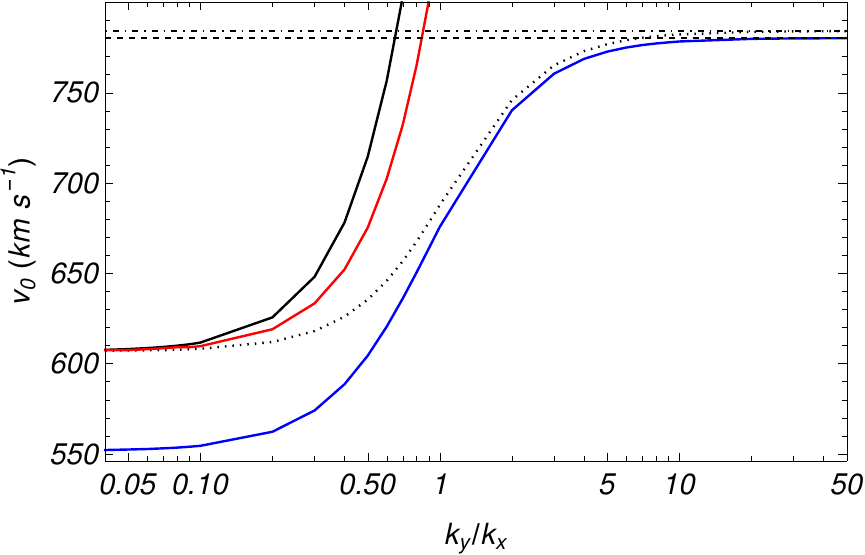}
    \caption{Critical flow velocity for the KHi onset as a function of $k_{y}/k_{x}$ in the ideal (dotted black line) and ambipolar (blue line) cases. The discontinuous horizontal lines represent the analytic critical velocities in the limit of nearly perpendicular propagation given in Equations~(\ref{eq:vcritb}) and (\ref{eq:vcritin}). The black line corresponds to the velocity for compressive stabilization in the ideal case. The red line represents the  flow velocity corresponding to  the maximum growth rate.}
    \label{fig:vcrit}
\end{figure}

For slow flows below the critical one, the ambipolar cases provide similar results concerning the fast surface waves compared to the ideal case. The main difference is that ${\rm Re}(s)$ has a small negative value in the ambipolar case, which is related to the  ambipolar damping, while in the ideal case ${\rm Re}(s) = 0$. Since the ambipolar damping is  weak when $k_x \ll k_{x}^{\rm cut.}$, this  is not appreciable at the scale of Figure~\ref{fig:betakh}.  When the critical flow velocity is reached, the FFSW and the BFSW converge and the stable  and unstable  branches appear in a similar fashion as happened in the incompressible limit (see the red and blue lines in the upper panel of Figure~\ref{fig:betakh}). The dependence of the critical flow velocity with the ratio $k_y/k_x$ is studied in Figure~\ref{fig:vcrit} for both ideal and ambipolar cases.  When $k_y/k_x \gg 1$, the ideal and ambipolar critical velocities consistently agree with those of Equations~(\ref{eq:vcritin})  and (\ref{eq:vcritb}), respectively. As $k_y/k_x$ decreases, both critical velocities decrease and the difference between them becomes larger. For $k_y/k_x \ll 1$, the critical velocity in the ambipolar case tends to the coronal Alfv\'en speed, while in the ideal case the critical velocity is the sum of the prominence and coronal Alfv\'en speeds, approximately. Thus, for small $k_y/k_x$ the effect of the ambipolar diffusion in decreasing the critical velocity is more important. 

Contrary to what happens in either the incompressible case or in the limit of nearly perpendicular propagation, the growth rate does not increase indefinitely with increasing $v_0$. This is due to the  effect of compressibility \citep[see details in][and references therein]{soler2012}. In the ideal case, the stable and unstable branches converge again for a second critical flow velocity and the interface becomes  stable once more for larger flow velocities, for which compressibility fully stabilizes the interface. For the model parameters, this happens at $v_0 \approx 960$~km~s$^{-1}$. The threshold velocity for compressive stabilization has been overplotted in Figure~\ref{fig:vcrit} to study its dependence with $k_y/k_x$. We find that when $k_y/k_x \gg 1$ the flow velocity for compressive stabilization tends to infinity, as consistent with the results in the incompressible limit. As $k_y/k_x$ gets smaller, so does the critical flow. It turns out that when $k_y/k_x \ll 1$ the critical flow velocity for the KHi onset and that for the compressive stabilization converge, so that there is no unstable range of flow velocities in the limit of purely parallel propagation. The flow velocity for which the instability growth rate reaches its maximum is also included in Figure~\ref{fig:vcrit}. This velocity represents the critical velocity for which the stabilizing effect of compressibility starts to dominate over the shear-induced instability.

\begin{figure}
    \centering
  \includegraphics[width=0.8\textwidth]{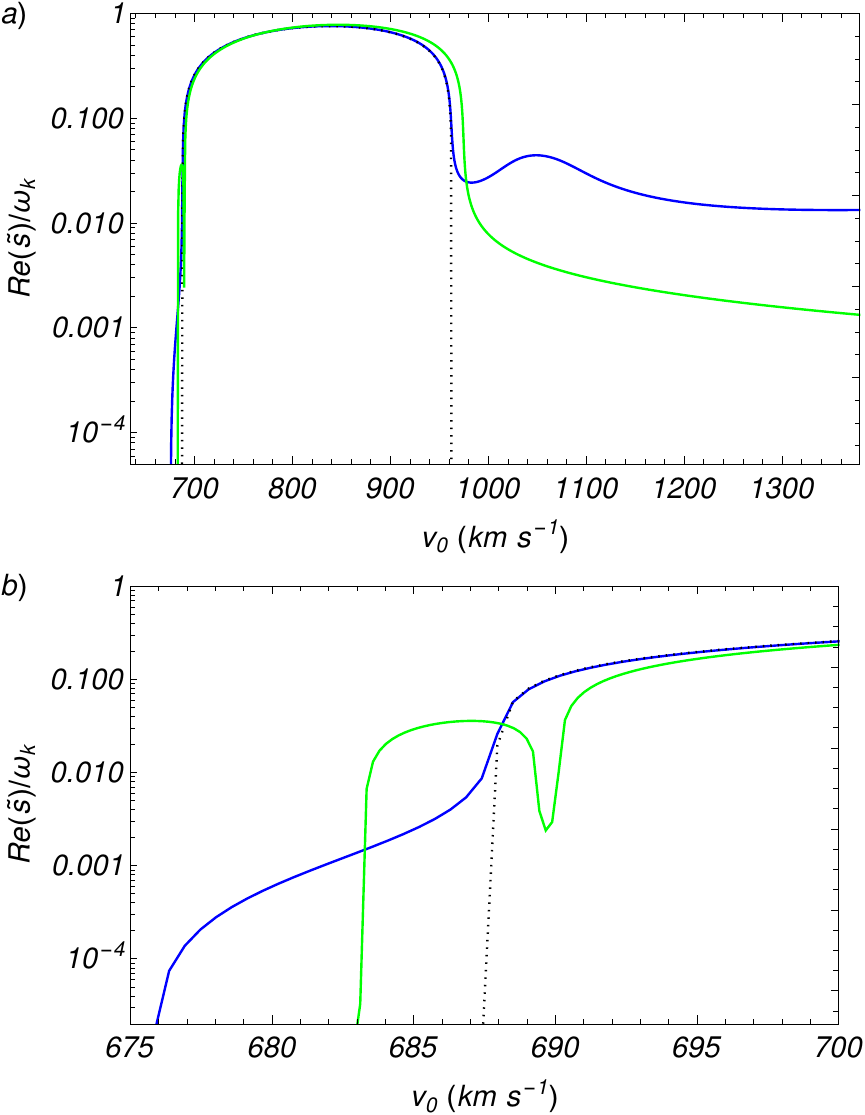}
   \caption{(a) Close-up view of the top panel of Figure~\ref{fig:betakh} around the range of flow velocities for which the KHi is excited. The blue line is the result in the pressuleress ambipolar case, while the discontinuous line is the pressuleress ideal result. For comparison purposes, the equivalent result in the complete case (with both gas pressure and ambipolar diffusion) is overplotted with a purple line. (b) Same results as in panel (a) but focusing on the initial part of the unstable region. Note that the vertical axis is in logarithmic scale in both panels.}
   \label{fig:zoom}
\end{figure}

The results in the ambipolar case initially follow a similar tendency as the ideal ones, but an important difference appears for large flow velocities. In the ambipolar case, complete stability does not happen when the compressibility-related critical flow velocity that stabilizes the ideal case is  reached. Instead,  ${\rm Re}(s)$ remains always positive although larger flow velocities are considered. This finding can be better visualized in Figure~\ref{fig:zoom}, which compares the ideal and ambipolar cases in a close-up view of the growth rate around the critical flow velocities discussed above. It is clear that ambipolar diffusion plays a destabilizing influence for large flow velocities that works against the stabilizing   role  of compressibility.

A remarkable feature seen in Figure~\ref{fig:zoom} is the small bump that the growth rate has for flow velocities slightly larger than the one that stabilizes the ideal case. This can be attributed to a coupling and avoided crossing between the FFSW and the FM (see the green line in Figure~\ref{fig:betakh}). This interaction between modes is absent from the ideal results and only happens in the presence of ambipolar diffusion. However, as discussed later in the complete case, this small bump might be an unwanted consequence of the pressureless approximation.

%So, it appears that the coupling with the FM is also important to understand why the unstable solution cannot  completely be stabilized in the ambipolar case.

\subsection{Complete case}

Once we have studied the incompressible limit and the compressible pressureless case, we tackle the investigation of the solutions of the full dispersion relation, including the effects of both compressibility and gas pressure. The fact that the gas pressure is now nonzero, and so the sound speed,  introduces a new type of magnetoacoustic mode into the scene. These are the slow magnetoacoustic surface waves, while in the pressureless approximation only the fast magnetoacoustic surface waves were present.

\begin{figure}
    \centering
    \includegraphics[width=0.8\linewidth]{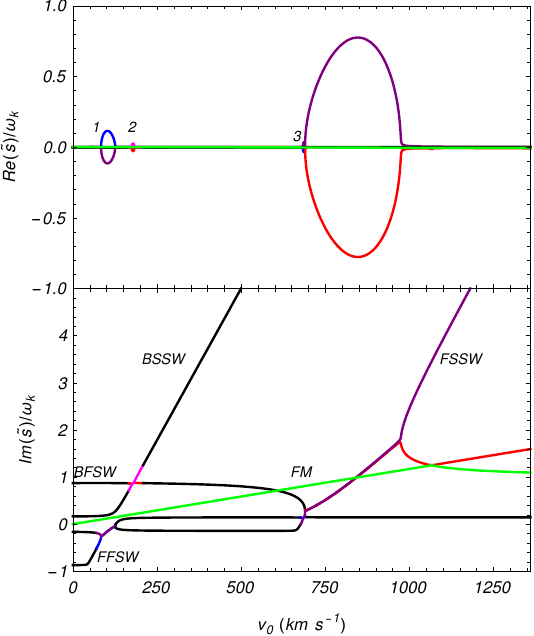}
    \caption{Same as Figure~\ref{fig:betakh} but for the complete case. The numbers 1, 2, and 3 in the upper panel denote the three instability regions discussed in the text.}
    \label{fig:complete}
\end{figure}

We plot Figure~\ref{fig:complete}  the real and imaginary parts of the growth rate as functions of the  flow velocity obtained in this complete case for the same parameters as those used before in the pressureless case of Figure~\ref{fig:betakh}. Now, the solutions  are the forward propagating fast surface wave (FFSW), the backward propagating fast surface wave (BFSW), the forward propagating slow surface wave (FSSW), and backward propagating slow surface wave (BSSW), in addition to the flow mode (FM).  The labels  in Figure~\ref{fig:complete} denote the mode character when $v_0 \to 0$, but the modes  interact and change character  in several occasions, which makes it difficult to track the individual modes according to their true properties when $v_0$ increases. An additional consequence  of this intricate interaction between the various modes is the appearance of three distinct regions of instability, which we label as 1, 2, and 3 according to their ordering with respect to the flow velocity.

The third region of instability (that for the largest flow velocities) is equivalent to that already present in Figure~\ref{fig:betakh} and is related to the interaction between the FFSW and BFSW. We have overplotted this region in Figure~\ref{fig:zoom}, in order to better compare the complete results with those in the pressuless approximation.  It is seen that the threshold velocity of the region in the complete case  is slightly larger than than of the pressureless ambipolar case, but still smaller than in the ideal result. On the other hand, we also find in the complete case that ambipolar diffusion removes the critical velocity of compressive  stabilization as in the pressureless case, although the growth rate is smaller in the complete case. In addition, we see that the small bump of the growth rate observed in the pressureless solution is absent from the complete results. To understand why this happens, we performed further calculations by progressively reducing the sound speed  (these results are not shown here). We found that the bump reappears and grows as   the sound speed decreases.  The inclusion of acoustic effects modifes somehow the way in which the various modes couple and  the coupling with the FM loses importance in the complete case. The situation is reminiscent of the mode couplings studied in Figure 4 of \citet{soler2009NA}. In the pressureless case, the coupling of the FM  with the surface waves resembles the ``anomalous coupling''  discussed in \citet{soler2009NA}, whereas in the complete case the coupling  is more like that in the ``weak coupling'' scenario.

\begin{figure}
    \centering
    \includegraphics[width=0.8\linewidth]{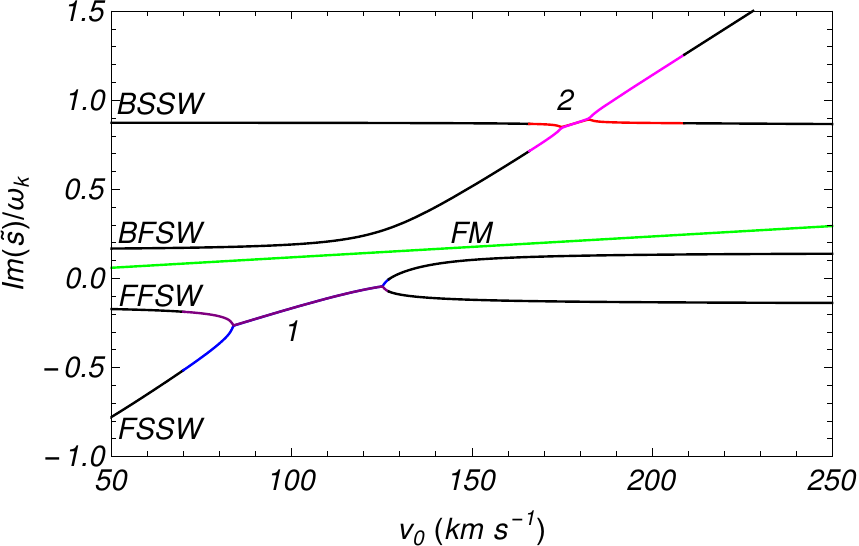}
    \caption{Close-up view of the lower panel of Figure~\ref{fig:complete} in the range of flow velocities where the first and second instability regions appear. The locations of these regions have been marked with the numbers 1 and 2, and are connected to the coupling and merging of different wave modes. The meaning of the other labels is the same as in Figure~\ref{fig:complete}.}
    \label{fig:zoom_im}
\end{figure}

Conversely, the first and second regions of instability seen in Figure~\ref{fig:complete} were not present in the pressureless results of Figure~\ref{fig:betakh} and appear at much smaller flow velocities. The first unstable region  appears at $v_{0} \approx 56$~km~s$^{-1}$ and extends up to $v_{0} \approx 127$~km~s$^{-1}$, while the second unstable region appears at $v_{0} \approx 166$~km~s$^{-1}$ and extends up to $v_{0} \approx 206$~km~s$^{-1}$. These velocities, specially the ones corresponding to the first region, are close to the typically observed flow velocities in prominences. The reason for the appearance of both first and second regions is related to mode couplings that involve the slow magnetoacoustic surface waves (see Figure~\ref{fig:zoom_im} for a close-up view of the couplings).

\begin{figure}
    \centering
    \includegraphics[width=0.8\linewidth]{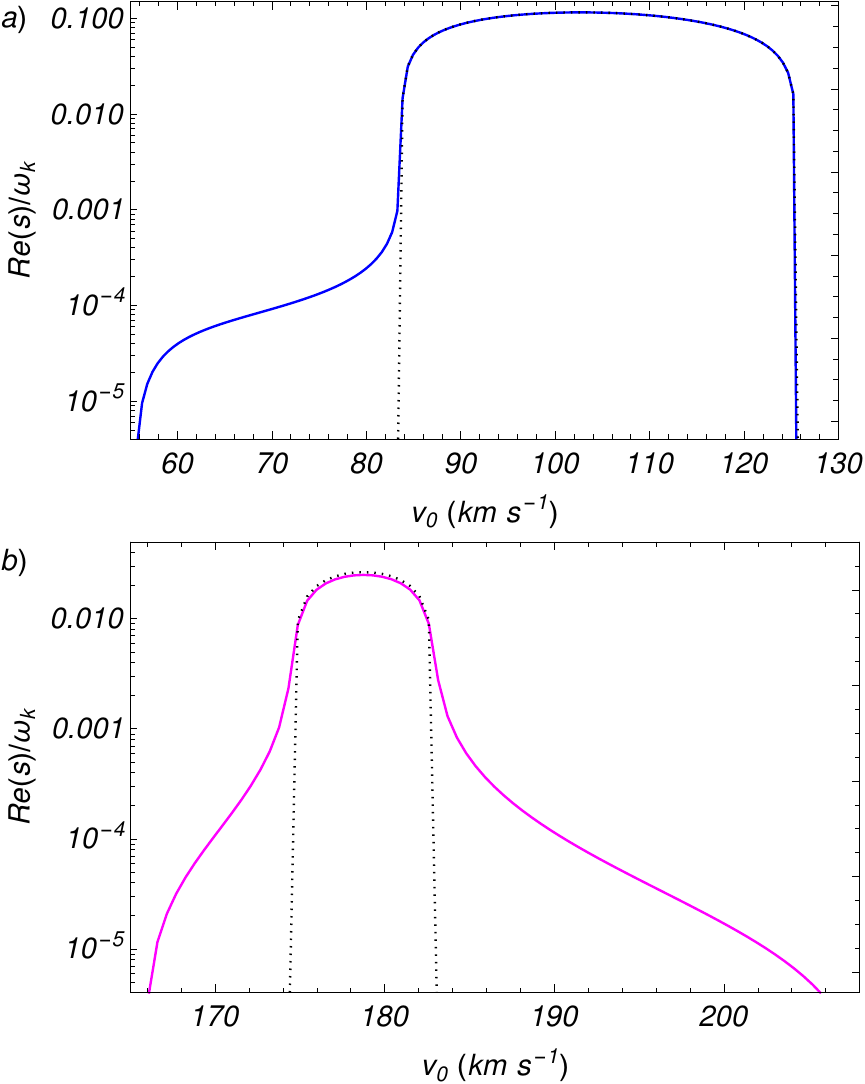}
    \caption{(a) Close-up view of the first instability region of Figure~\ref{fig:complete} corresponding to the complete case. The solid line is the  result in the  ambipolar case, while the discontinuous line is the  ideal result.  (b) Same  as in panel (a) but for the second instability region of Figure~\ref{fig:complete}. Note that the vertical axis is in logarithmic scale in both panels.}
    \label{fig:zoom_c}
\end{figure}

Detailed close-up views of these new instability regions are depicted in Figure~\ref{fig:zoom_c}, where the results in the ideal case are also plotted for comparison. We find that ambipolar diffusion has a pronounced effect on the flow velocities that limit each region. In the first region, we see that the threshold velocity gets reduced from $v_{0} \approx 83$~km~s$^{-1}$ in the ideal case to $v_{0} \approx 56$~km~s$^{-1}$ in the presence of ambipolar diffusion. Conversely, the critical velocity (associated to compressibility) that stabilizes the region is not affected by the ambipolar diffusion. Regarding the second region, we obtain that ambipolar diffusion shifts both initial and final velocities compared to those of the ideal case, so that the range of unstable velocities widens.

%In the upper panel are represented the first forward fast-slow coupling and the backward fast-slow coupling, while the lower panel shows the second forward fast-slow coupling. In the upper panel, the effect of the ambipolar diffusion is almost negligible in the first instability region, as the difference between the ideal and the ambipolar cases for both modes are almost identical if we compare them. However for the second region the differences are more visible, since the unstable mode has a slightly smaller value compared to the ideal case, while the damped mode has a small increase. This effect is related with the damping produced by the ambipolar diffusion coefficient. In the lower panel the transition between the third instability region and the main region is seen clearly as both region are connected smoothly whereas in the ideal case the third region is closed and the main region appears without interacting. In both plots the effect of the ambipolar diffusion is seen as the real part of the growth rate has always a non-zero value, contrary to the ideal case that when there is a non-unstable region the growth rate is purely imaginary.

\begin{figure}
    \centering
    \includegraphics[width=0.8\linewidth]{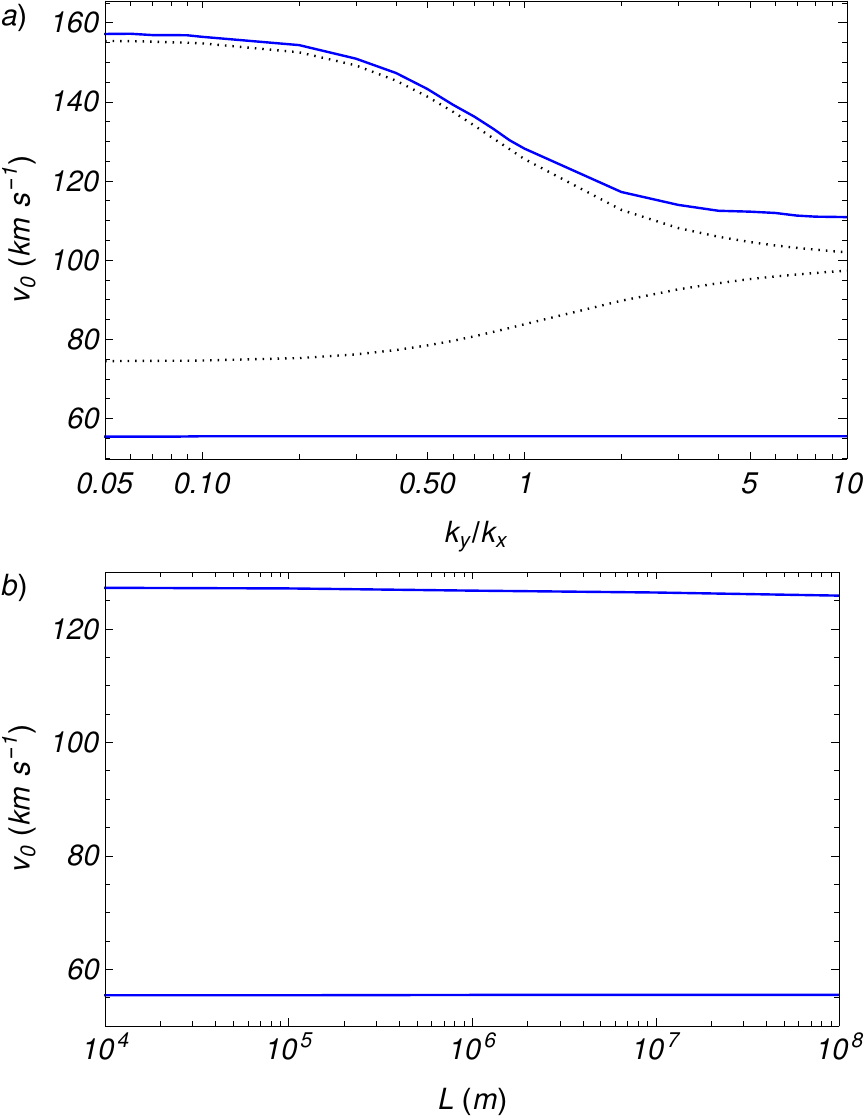}
    \caption{First unstable region in the complete case. (a) Range of unstable flow velocities for the KHi onset  as a function of $k_{y}/k_{x}$ in the presence of ambipolar diffusion (between the blue solid lines) and in the ideal case (between the black dotted lines). (b) Same as (a) but as a function of the normalization length scale, $L$, for $k_{y}/k_{x}=1$. Only the results in the presence of ambipolar diffusion are displayed in (b).}
    \label{fig:vcrit_c}
\end{figure}

In connection  to the observed flow velocities in solar prominences, it is clear that the third unstable region occurs for much larger velocities than those observed. The second unstable region takes place is smaller flow velocities, but still somewhat larger than the observed ones. Therefore, neither the second nor the third region of instability are able to explain the observations of KHi triggering in prominences. However, the flow velocities corresponding to the first region are much closer to the observed values. The fact that the ambipolar diffusion further reduces the threshold velocity of this region increases its potential importance. We have studied how the velocity range of the first region of instability changes with the ratio $k_y/k_x$. This is displayed in Figure \ref{fig:vcrit_c}a. The threshold flow velocity is independent of $k_y/k_x$, while the maximum unstable velocity in this region decreases/increases for increasing/decreasing $k_{y}/k_x$. For comparison, we have also represented the critical velocities in the ideal case, which shows that the unstable region gets narrower as $k_{y}/k_x$ increases and disappears when $k_{y}/k_x \to \infty$. Conversely, in the ambipolar case this unstable region exists for all values of $k_{y}/k_x$.

%In this plot are represented both the critical flow and the maximum velocity where the KHi is present and the results obtained for the ambipolar case have been compared with the ideal case. The effect of the ambipolar diffusion on the critical flow velocity produces a fixed critical mass flow independent of the wavenumber, which remains in the observational range, however the maximum flow for this region decreases for increasing $k_{y}$, having a slightly larger value compared to the ideal case. The difference between the maximum flow of the ideal and ambipolar cases widens when the incompressible limit is reached.

Finally, we have also studied the effect of the normalization lenghtscale, $L$, on the first unstable region. When $L$ varies, the importance of the ambipolar diffusion is modified, being more significant for small scales than for large scales. Figure \ref{fig:vcrit_c}b shows the effect of the lenghtscale on the extent of the first unstable region when  $k_{y}/k_{x}=1$.  The maximum velocity has a smooth decrease towards the ideal maximum velocity when $L$ increases, but since the maximum flow in the ambipolar case is already close to the one in the ideal case, the change is small. Conversely, the threshold flow velocity remains virtually constant for all studied values of $L$, although the ambipolar diffusion becomes very weak for large $L$. As already discussed in the pressureless case, the reason for the independence of the threshold flow on the efficiency of ambipolar diffusion can be attributed to the model set-up made of an abrupt boundary. If a continuous boundary was considered, or if there were dissipation in the coronal region, the threshold flow velocity should tend to the ideal value as $L$ increases.

%Next we study the effect of the lenghtscale in the first instability region. The main effect of the variation of the lengthscale is the relevance of the ambipolar diffusion. For larger values the ambipolar diffusion tends to values that are almost negligible and the ideal case can be recovered, while for smaller values the ambipolar diffusion takes a more prominent role. Figure \ref{fig:vcrit_c}b shows the effect of the lenghtscale for a fixed ratio of $k_{y}/k_{x}=1$ in the critical and maximum flows of the first instability region. The critical flow remains almost uniform for all values of $L$ studied, even for the one that approach the ideal case. This result can be attributed to the set-up of this work where there is an abrupt boundary and there is no dissipation in the coronal region. If there was a diffusion term in the corona there would be a smooth increase towards the ideal critical flow. On the other side, the maximum flow has a smooth decrease towards the ideal maximum flow, however the difference between the maximum flow of the ambipolar case is already close to the one in the ideal case and thus the change is small.

% \begin{figure}
%     \centering
%     \includegraphics[width=0.8\linewidth]{comparison.eps}
%     \caption{Real part of the growth rate for the pressureless case and the main instability region of the complete case as function of $v_{0}/v_{\rm Ap}$ evaluated for $k_{x}L=k_{y}L=\pi$. \\ 
%     {\bf AQUESTA FIGURA ES POT LLEVAR PERQUE ES JUNTA AMB LA FIGURA~5}}
%     \label{fig:comparison}
% \end{figure}

% Figure \ref{fig:comparison} compares the main instability region for both the pressureless, represented with the blue line for the unstable mode and black line for the stable mode respectively, and the complete case, represented with the red line for the unstable mode and the magenta line for the stable mode respectively. The effect of compressibility is seen as both unstable region have an almost closed unstable region, even though if it is not completely closed due to the presence of the ambipolar diffusion. Both cases provide similar values with 

\section{Conclusions}

In this paper, we have studied the onset of the KHi due to a shear flow in an interface with prominence properties on one side and coronal properties on the other side. We have considered the effect of ambipolar diffusion in the partially ionized prominence plasma. We have obtained a general dispersion relation for the linear perturbations on the interface, which includes the effects of plasma compressibility, gas pressure, and ambipolar diffusion. Before studying this complete case, we have analyzed some limit cases, namely the incompressible limit and the compressible presureless case.

In the incompressible limit, we found that the ambipolar diffusion has no effect and the KHi happens for super-Alfv\'enic flows alone as in the ideal case. In the absence of compressibility, the range of unstable flow velocties is unlimited once the threshold value is exceeded.  Conversely, in the compressible pressureless case the ambipolar diffusion produces two interesting results. Firstly, the critical velocity for the KHi onset is slightly reduced, even though it remains super-Alfv\'enic and does not reach values consistent with the observed flows in prominences. Secondly, the ambipolar diffusion has a destabilizing influence that works against the attempt of compressibility  to suppress the KHi for large flow velocities.

In the general case, we essentially recover the results of the compressible presureless case, but some important differences are found. The inclusion of the acoustic effects produces the appearance of additional regions of instability that happen for slower flows, which are associated with couplings with the slow magnetoacoustic waves. The ambipolar diffusion has  the effect of widening these new regions of instability. In particular, the unstable flow velocities corresponding to the first region of instability approach similar sub-Alfv\'enic values as those reported in the observations.

To end with, it is fair to mention that  we neglected some effects like gravity, dissipation in the coronal region, and other non-ideal terms for the sake of simplicity. The inclusion of these effects in future works could provide a more realistic investigation of the KHi onset. Another further study could be comparing the results obtained here in the single-fluid approximation with those of the two fluid approach used in  \citet{soler2012} and \citet{martinez2015}, although they neglected the effect of helium,  which was considered here in the computation of the ambipolar diffusion.

\begin{acks}
 The authors thank the unknown referee for constructive comments that helped to improve the paper.
\end{acks}

\begin{authorcontribution}
The two authors contributed to the study conception and design. Analytic derivations, computations, analysis of results, and graphical representations were performed by LM under the supervision of RS. The first draft of the manuscript was written by LM, with later additions and corrections by RS. The two authors read and approved the final manuscript.
\end{authorcontribution}

\begin{fundinginformation}
This publication is part of the R+D+i project PID2023-147708NB-I00, funded by MCIN/AEI/10.13039/501100011033 and by FEDER, EU. LM is supported by the predoctoral fellowship FPI\_002\_2022  funded by CAIB.
\end{fundinginformation}

\begin{ethics}
\begin{conflict}
The authors declare no conflicts of interest.
\end{conflict}
\end{ethics}

\begin{appendix}
    \section{Dispersion relation in the complete case}
    \label{app}
    In the complete case, the dispersion relation is $\mathcal{D}(s)=0$, where $\mathcal{D}(s)$ is the dispersion function that can be computed from the following determinant,
    \begin{eqnarray}
    \mathcal{D}(s)=
    \begin{vmatrix}
        a_{11} & a_{12} & a_{13} \\
        a_{21} & a_{22} & a_{23} \\
        a_{31} & a_{32} & a_{33} \\
    \end{vmatrix},
    \end{eqnarray}
where
\begin{align}
    a_{11} &= \frac{k_{\rm c}\left( c_{\rm sc}^{2}v_{\rm Ac}^{2}k_{x}^{2}+\tilde{s}^{2}\left( c_{\rm sc}^{2}+v_{\rm Ac}^{2}\right)\right)}{\tilde{s}^{3}\left( \tilde{s}^{2} +k_{x}^{2}v_{\rm Ac}^{2}\right)},\\
    a_{12} &= \frac{k_{\rm p+}\left( \tilde{s}^{2}\tilde{v}_{\rm Ap}^{2}+c_{\rm sp}^{2}\left( \tilde{s}^{2}-(k_{\perp}^{2}-k_{\rm p+}^{2})v_{\rm Ap}^{2} + (k_{\perp}^{2}+k_{x}^{2}-k_{\rm p+}^{2})\tilde{v}_{\rm Ap}^{2}\right) \right)}{\tilde{s}^{3}\left( \tilde{s}^{2} +k_{x}^{2}\tilde{v}_{\rm Ap}^{2}\right)}, \\
    a_{13} &= \frac{k_{\rm p-}\left( \tilde{s}^{2}\tilde{v}_{\rm Ap}^{2}+c_{\rm sp}^{2}\left( \tilde{s}^{2}-(k_{\perp}^{2}-k_{\rm p-}^{2})v_{\rm Ap}^{2} + (k_{\perp}^{2}+k_{x}^{2}-k_{\rm p-}^{2})\tilde{v}_{\rm Ap}^{2}\right) \right)}{\tilde{s}^{3}\left( \tilde{s}^{2} +k_{x}^{2}\tilde{v}_{\rm Ap}^{2}\right)},\\
    a_{21} &= \frac{\rho_{\rm c}\left( c_{\rm sc}^{2}v_{\rm Ac}^{2}k_{x}^{2}+\tilde{s}^{2}\left( c_{\rm sc}^{2}+v_{\rm Ac}^{2}\right)\right)}{\tilde{s}^{3}},\\
    a_{22} &= -\frac{\rho_{\rm p}\left(\tilde{s}^{2}\tilde{v}_{\rm Ap}^{2}+c_{\rm sp}^{2}\left( \tilde{s}^{2}+v_{\rm Ap}^{2}(k_{\rm p+}^{2}-k_{y}^{2}) +\tilde{v}_{\rm Ap}^{2}(k_{\perp}^{2}-k_{\rm p+}^{2})\right) \right)}{\tilde{s}^{3}},\\
    a_{23} &= -\frac{\rho_{\rm p}\left(\tilde{s}^{2}\tilde{v}_{\rm Ap}^{2}+c_{\rm sp}^{2}\left( \tilde{s}^{2}+v_{\rm Ap}^{2}(k_{\rm p-}^{2}-k_{y}^{2}) +\tilde{v}_{\rm Ap}^{2}(k_{\perp}^{2}-k_{\rm p-}^{2})\right) \right)}{\tilde{s}^{3}},\\
    a_{31} &= i k_{x}B_{0} \frac{k_{\rm c}\left( c_{\rm sc}^{2}v_{\rm Ac}^{2}k_{x}^{2}+\tilde{s}^{2}\left( c_{\rm sc}^{2}+v_{\rm Ac}^{2}\right)\right)}{\tilde{s}^{3}\left( \tilde{s}^{2} +k_{x}^{2}v_{\rm Ac}^{2}\right)}, \\
    a_{32} &= \frac{i k_{x}B_{0} k_{\rm p+}\left(\tilde{s}^{2}\tilde{v}_{\rm Ap}^{4}+c_{\rm sp}^{2}\left( \tilde{s}^{2}v_{\rm Ap}^{2}+(k_{\rm p+}^{2}-k_{y}^{2})v_{\rm Ap}^{2}\tilde{v}_{\rm Ap}^{2}+\tilde{v}_{\rm Ap}^{4}(k_{\perp}^{2}-k_{\rm p+}^{2})\right) \right)}{\tilde{s}^{3}v_{\rm Ap}^{2}\left( \tilde{s}^{2} +k_{x}^{2}\tilde{v}_{\rm Ap}^{2}\right)},\\
    a_{33} &= \frac{i k_{x}B_{0} k_{\rm p-}\left(\tilde{s}^{2}\tilde{v}_{\rm Ap}^{4}+c_{\rm sp}^{2}\left( \tilde{s}^{2}v_{\rm Ap}^{2}+(k_{\rm p-}^{2}-k_{y}^{2})v_{\rm Ap}^{2}\tilde{v}_{\rm Ap}^{2}+\tilde{v}_{\rm Ap}^{4}(k_{\perp}^{2}-k_{\rm p-}^{2})\right) \right)}{\tilde{s}^{3}v_{\rm Ap}^{2}\left( \tilde{s}^{2} +k_{x}^{2}\tilde{v}_{\rm Ap}^{2}\right)}.
\end{align}

In the incompressible and pressureless limits, the dispersion relation reduces to those of Equations~(\ref{eq:incompressible}) and (\ref{eq:beta0}), respectively.

\end{appendix}

     % format of references provided by the journal (.bst)
\bibliographystyle{spr-mp-sola}
     % name your Bibtex file containing your references (.bib)
\bibliography{biblio}  

     % Checking: look if the file containing the ``\bibitem'' exits
     %           so check if the .bbl file exist (bibTeX compilation)
%\newpage     

\end{document}